\documentclass[12pt,draftcls,onecolumn]{IEEEtran}
\usepackage{upgreek}
\usepackage{graphicx}
\usepackage{amsmath}
\usepackage{multicol}
\usepackage{multirow}
\usepackage{stfloats}
\usepackage{booktabs}
\usepackage{mathrsfs}
\usepackage{enumerate}
\usepackage{amssymb}
\usepackage{algorithm}
\usepackage{algpseudocode}

\usepackage{array}
\usepackage{underscore}
\usepackage{url}

\usepackage{hyperref}
\hypersetup{hidelinks}
\usepackage{flushend}

\usepackage[square, comma, numbers, sort&compress]{natbib}
\usepackage{graphicx}
\usepackage{epstopdf}
\usepackage{caption}
\usepackage{diagbox}

\usepackage{subfigure}
\usepackage{color}

\usepackage[outerbars,color]{changebar}
\ifx\pdfoutput\undefined
\else\ifnum\pdfoutput>0
\usepackage{pdfcolmk}
\fi\fi
\cbcolor{red}
\usepackage{fancyhdr}

\makeatletter
\newcommand{\rmnum}[1]{\expandafter\@slowromancap\romannumeral #1@}
\makeatother

\usepackage{array}  \newcommand{\PreserveBackslash}[1]{\let\temp=\\#1\let\\=\temp}  \newcolumntype{C}[1]{>{\PreserveBackslash\centering}p{#1}}  \newcolumntype{R}[1]{>{\PreserveBackslash\raggedleft}p{#1}}  \newcolumntype{L}[1]{>{\PreserveBackslash\raggedright}p{#1}}

\ifCLASSINFOpdf
\else
\fi

\begin{document}

\title{\huge Resonant Beam Communications with Photovoltaic Receiver for Optical Data and Power Transfer}

\author{Mingliang~Xiong,
        Qingwen~Liu*, Mingqing Liu,  Xin Wang, and Hao Deng 

\thanks{* The corresponding author: Qingwen Liu}

\thanks{The material in this paper has been presented in part to the 53rd IEEE International Conference on Communications (ICC), Shanghai, China, May 20-24, 2019.}

\thanks{
	M. Xiong, Q. Liu, M. Liu, and Hao Deng
	are with the College of Electronics and Information Engineering, Tongji University, Shanghai, China
	(email: xiongml@tongji.edu.cn,
	qliu@tongji.edu.cn, clare@tongji.edu.cn,  dashena85754@qq.com).}
\thanks{
	X. Wang is with the Key Laboratory for Information Science of Electromagnetic
	Waves (MoE), the Shanghai Institute for Advanced Communication and
	Data Science, the Department of Communication Science and Engineering,
	Fudan University, Shanghai, China (e-mail: xwang11@fudan.edu.cn).}
}
\maketitle

\begin{abstract}
The vision and requirements of the sixth generation (6G) mobile communication systems are expected to adopt freespace optical communication (FSO) and wireless power transfer (WPT). The laser-based WPT or wireless information transfer (WIT) usually faces the challenges of mobility and safety. We present a mobile and safe resonant beam communication
(RBCom) system, which can realize high-rate simultaneous wireless information and power transfer (SWIPT). We propose an 
analytical model to depict its carrier beam and information transfer procedures. The numerical
results show that RBCom can achieve more than $40$~mW charging power and $1.6$ Gbit/s channel capacity with orthogonal frequency division multiplexing (OFDM) scheme, which can be applied in future scenario where power and high-rate data are simultaneously desired.
\end{abstract}

\begin{IEEEkeywords}
Resonant beam system, free space laser resonator, laser communications, simultaneous wireless information and power transfer.
\end{IEEEkeywords}

\IEEEpeerreviewmaketitle

\section{Introduction}
\subsection{Motivation}

 The vision and requirements of the sixth generation
 (6G) mobile communication systems are presented to include
 FSO and wireless charging for
 mobile devices~\cite{a180820.09,a181126.01}, which will enhance the ability of computing and communication of mobile Internet of things (IoT) devices and promote the development of big data and intelligent cognition over  IoT~\cite{a181126.02,a181126.04,a181126.03,a181126.05,a181126.06}. The laser based
simultaneous wireless information and power transfer
(SWIPT) technologies can achieve high bit-rate and high power. However, they
usually face the challenges of mobility and safety.

The resonant beam communications (RBCom) system proposed here is an optical wireless data and power transfer system. The novelty is as follows. In contrast to conventional established charging mechanisms such as inductive coupling and magnetic resonance coupling, the resonant beam system features high-power safety, long range, and non-mechanical mobility. Since the RBCom system generates directed resonant beam to the receiver, most energy is focused on the receiver. However, conventional energy harvesting (EH) from radio frequency (RF) radiation can only obtain a small portion of the generated power, as RF radiation has a large divergence angle. On the other hand, the RBCom is a free-space optical communication (FSO) system. With the development of mobile Internet, the increasingly demand for ultrahigh bit-rate and low-delay mobile communications has been put forward. As an available information carrier, light is capable of providing ultra-wide bandwidth. Therefore, FSO has a broad application prospect in indoor scenario in the future. 

\begin{figure}
	\centering
	\includegraphics[width=4.3in]{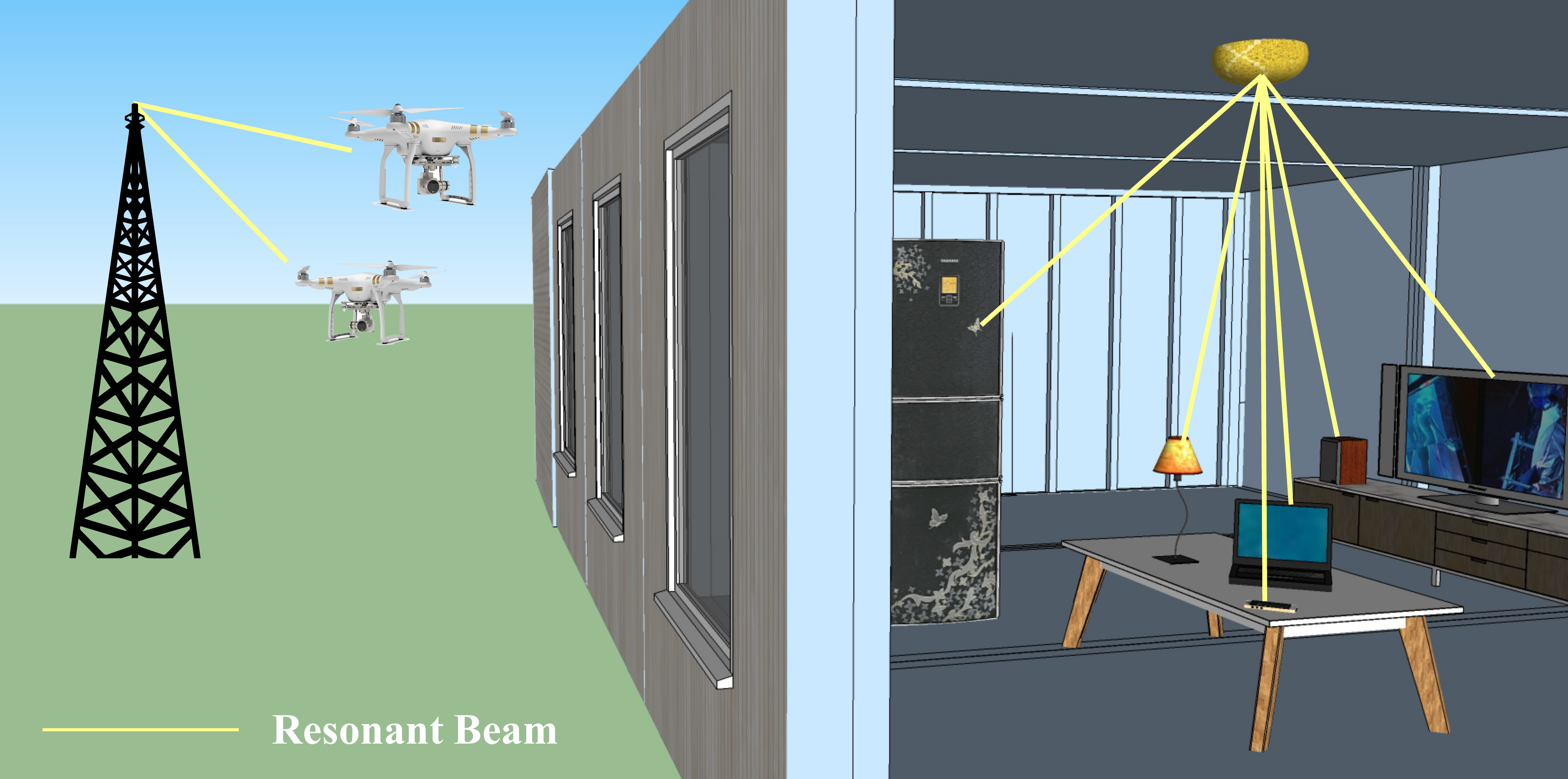}
	\caption{Data and power transfer via  resonant beam}
	\label{fig:scenario}
\end{figure}

As shown in Fig.~\ref{fig:scenario}, the RBCom system is designed for mobile data and power transfer, which connects the transmitter and the receiver with an optical beam link. Two retroreflectors are embedded in the transmitter and receiver, respectively, enabling non-mechanical mobility. Photons resonate between the transmitter and the receiver, forming a resonant beam which plays the role of information carrier. The detail mechanism is presented  in Section~\ref{sec:bkg}.

\subsection{Literature Review}

The design concept of RBCom origins from the design of the very long laser proposed by G. J. Linford \emph{et al.} in 1973 \cite{a190318.02,a190318.01}. Laser cavity length up to $30~\mbox{km}$ was studied, and the application of atmospheric pollution detection was discussed. In order to reduce the impact of misalignment of two plain mirrors in a long laser cavity, two retro-reflectors were used to substitute for the plain mirrors. Therefore, Linford's long laser has the features of self-pointing with a varying degree, as well as the property of fail-safe self-terminating oscillation if the laser beam is interrupted.

In 2016, Q.~Liu \emph{et al.} introduced the RBC technology, based on the very long laser design, to charge mobile devices via the retro-reflected resonant beam between the transmitter and the receiver~\cite{a180727.01}. Afterwards, Q. Zhang \emph{et al.} analyzed the transmission model of the RBC system, and then, proposed the adaptive RBC (ARBC) system to improve the energy utilization~\cite{a180820.03,a180820.04}. In 2018, W. Fang \emph{et al.} proposed the fair scheduling algorithm in RBC system~\cite{a180820.05}, and M. Xiong \emph{et al.} proposed the time-division multiple access (TDMA) design in the RBC system to enable the multi-user charging control~\cite{a180820.06}. These works on RBC provide the significant basis to the research of RBCom.

\subsection{Contribution}	
We at first present the resonant beam communication (RBCom) system design. Then, we propose the analytical model of the RBCom system to depict its power transfer and information transfer procedures, respectively. Relying on the numerical evaluation, we analyze the operating point, frequency response, capacity, and output charging power of the RBCom system. We find that the RBCom system can achieve more than 40 mW charging power and $1.6$ Gbit/s channel capacity with orthogonal frequency division multiplexing (OFDM) scheme.

The rest of the paper is organized as follows. The design of the RBCom system and its mechanism are presented in Section~\ref{sec:bkg}. The analytical model of the RBCom system is presented in Section~\ref{sec:ana}. The numerical results are presented in Section~\ref{sec:res}. Finally,  conclusions are drawn in Section~\ref{sec:con}.

\section{System Design}
\label{sec:bkg}

\begin{figure}
	\centering
	\includegraphics[width=4.3in]{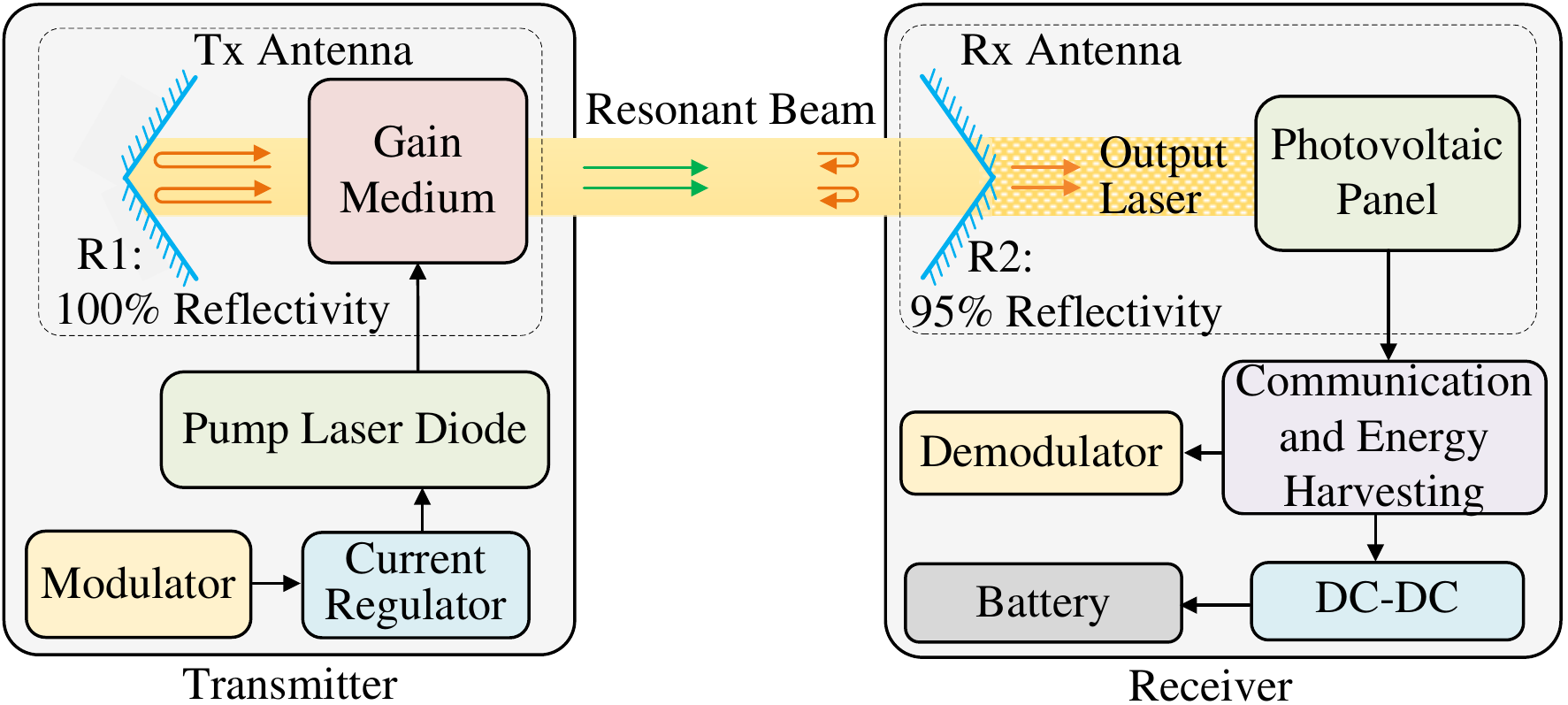}
	\caption{Resonant beam communication system diagram}
	\label{fig:RBC}
\end{figure}

Figure~\ref{fig:RBC} depicts the RBCom system diagram relying on the RBC power transfer presented in~\cite{a180820.03,rbcom2019}, the light beam resonates between the two retro-reflectors R1 and R2, while stimulating the gain medium for optical amplification. Therefore, RBCom has two key features:
\begin{itemize}
	\item {\it Mobility}: The RBCom system cavity comprises of two retroreflectors (one in the transmitter, and the other in the receiver) and a gain medium between the retroreflectors. Because the retroreflector reflects the incident radiation back to its source, photons generated by the gain medium can oscillate between the transmitter and the receiver regardless of the position of the receiver. The oscillating photons are amplified by the gain medium, forming the resonant beam that connects the transmitter and the receiver.
	\item{\it High-power Safety}: Any obstacles blocking the resonant beam path will terminate the resonance immediately. In this case, the power transfer procedure can be terminated, avoiding the harm to obstacles. As the obstacle entering the transmission path and approaching the beam center, diffraction loss in the cavity increases. According to laser cavity theory, if the diffraction loss becomes larger than the system gain produced by the gain medium, the resonance will cease in a very short time.  Inversely, as the obstacle leaving the transmission path, the intra-cavity loss becomes smaller than the small-signal gain factor, so the resonance is built up again automatically.
\end{itemize}

 	R2 allows parts of photons passing through to form a laser beam. The photovoltaic (PV) panel converts the coupled output laser into electricity. At last, the battery can be charged with the designed output voltage or current via the direct current to direct current (DC-DC) converter. The resonant beam system has the ability of charging batteries with sufficient power. As reported in \cite{a180820.07}, the charging power at the receiver ($1.675$~m from the transmitter) is up to $2$~W. With this power level, many consumer electronics such as smart watch and smart bracelet can be charged. In the future, the transmission distance and power can be improved by decreasing the intra-cavity loss. In addition, the charging power can be further increased by improving the electro-optic conversion efficiency and photoelectric conversion efficiency. The information can be modulated on the resonant beam, so that an optical link for data and power transfer can be created.

As shown in Fig.~\ref{fig:RBC}, the RBCom system comprises the transmitting (Tx) antenna and the receiving (Rx) antenna which are the primary parts of the RBC system. Besides, the system integrates a modulator and a linear current regulator at the transmitter. At the same time, the system includes a shunt circuit for communication and energy harvesting (CEH) at the receiver. As the input current of the pump laser is modulated by the modulator, the resonant beam as well as the output laser carries not only the power but also the information. The CEH module receives the output current of the PV panel, and then, separates the signal to the communication branch and guides the bias current to the energy harvesting branch. 

\section{Analytical Model}
\label{sec:ana}

\begin{figure*}
	\centering
	\includegraphics[width=6.5in]{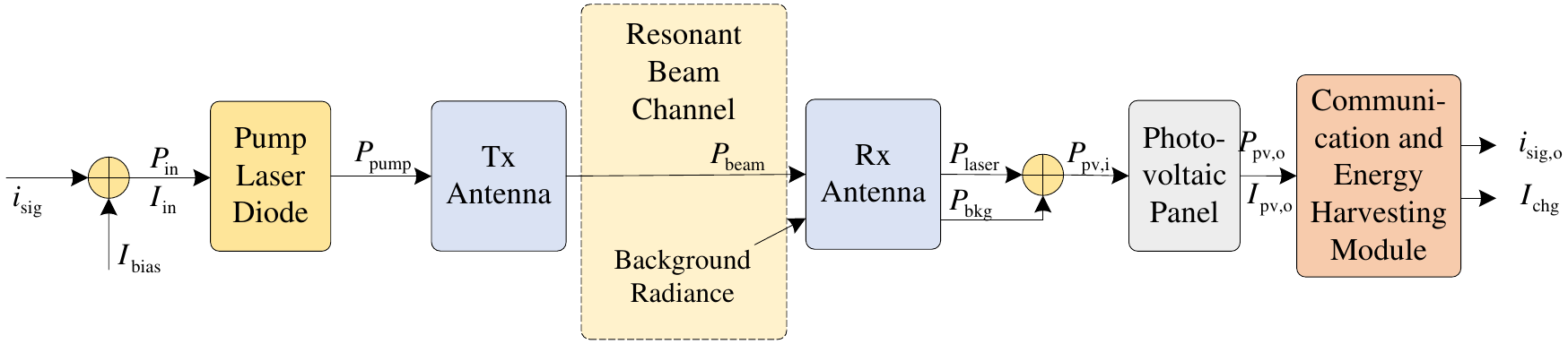}
	\caption{Analytical model for resonant beam communication system}
	\label{fig:AlyModel}
\end{figure*}

 Figure~\ref{fig:AlyModel} depicts the transmission model of the RBCom system. The input current  $I_{\rm in}$ of the pump LD generated by the current regulator comprises the signal current $i_{\rm sig}$ and the direct-current (DC) bias current $I_{\rm bias}$. The power of the background light which has passed through the Rx antenna is $P_{\rm bkg}$. The output power and current from the PV panel are $P_{\rm pv,o}$ and $I_{\rm pv,o}$, respectively. The CEH module separates $I_{\rm pv,o}$ into signal current $i_{\rm sig,o}$  and  bias current $I_{\rm chg}$.
	
\subsection{Power Transfer}
	The power transfer procedure in the RBCom system includes 4 stages: 1) the input power $P_{\rm in}$ to the pump laser power $P_{\rm pump}$; 2) $P_{\rm pump}$ to the resonant beam power $P_{\rm beam}$; 3) the resonant beam is partially outputted by Rx antenna, forming a laser whose power is $P_{\rm laser}$; and, 4) $P_{\rm laser}$ to the output electrical power $P_{\rm pv,o}$. On analysis, we find the linear relationship between the input current and the output current of this system.
	
\subsubsection{Pump LD}

	The pump LD generates a laser beam to stimulate and drive the gain medium. The current-power (I-P) characteristic of the pump LD can be described as~\cite{a180720.01}:
\begin{equation}
	P_{\rm pump}(I_{\rm in})=\frac{hc}{q\lambda}\eta_{\rm e}[I_{\rm in}-I_{\rm th}],
	\label{equ:P-I-LD}
\end{equation}
and
\begin{equation}
	\eta_{\rm e}=\eta_{\rm inj}\frac{\gamma_{\rm out}}{\gamma_{\rm c}},
\end{equation}
	where $h$ is Planck's constant; $c$ is the speed of light in vacuum; $q$ is the electron charge; $\lambda$ is the emission wavelength; $\eta_{\rm e}$ is the external quantum
	efficiency; and  $I_{\rm th}$ is the constant threshold current. The carrier injection efficiency $\eta_{\rm inj}$ and $I_{\rm th}$ are temperature-dependent parameters.  $\gamma_{\rm out}/\gamma_{\rm c}$ is the photon extraction efficiency. We can find that the relationship between the pump laser power and the input current above the threshold is close to linearity.
	
%

\subsubsection{Coupled Output Laser}
	The combination of the Tx and Rx antennas are similar to that of a LD pumped solid state laser. For simplicity, the RBCom system can be modelled as a long-cavity solid-state laser. As analyzed in~\cite{a180820.07}, the output laser power from the Rx antenna can be described as:

\begin{equation}
	P_{\rm laser}=\eta_{\rm s}P_{\rm pump}f(d)+C,
	\label{equ:P-laser}
\end{equation}
	where $\eta_{\rm s}$ is the conversion efficiency of population inversion; $f(d)$ is the function determined by the distance $d$ between the two antennas; and $C$ is a constant value depending on the internal cavity parameters. The gain medium absorbs the power from the incident pump laser to realize  population inversion, which is a necessary step for generating the laser beam. The function $f(d)$ can be described as~\cite{a180820.07}:
\begin{equation}
	f(d)=\dfrac{2(1-R)p}{(1+R){\rm e}^{\frac{-2\uppi a^2}{{\lambda}d}}-(1+R)\ln{R}},
	\label{equ:fd}
\end{equation}
where $R$ is the reflectivity of R2, $p$ is the overlap efficiency, and $a$ is the radius of the aperture that the beam passes through. In summary, the above procedure exhibits a linear relationship between $P_{\rm laser}$ and $P_{\rm pump}$ at a certain $d$.

\subsubsection{PV Panel}
\begin{figure}
	\centering
	\includegraphics[width=3.5in]{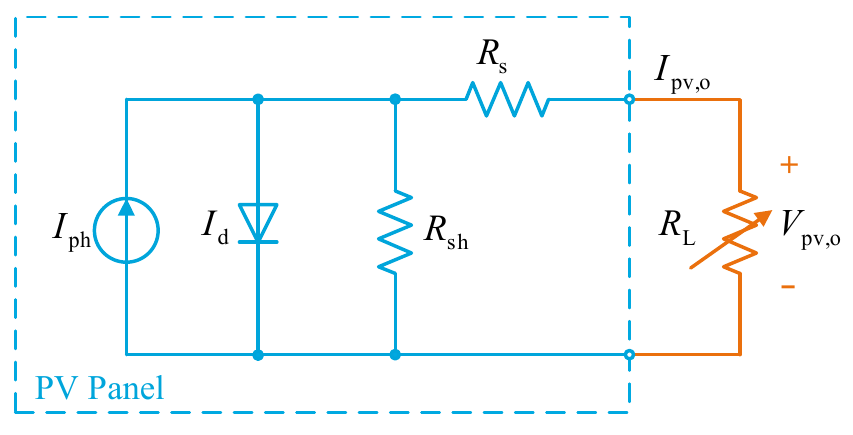}
	\caption{Receiver DC equivalent circuit}
	\label{fig:PV-equ}
\end{figure}
The PV panel converts the coupled output laser into electrical power. An ideal PV panel can be modelled as a photo-current source $I_{\rm ph}$ connected in parallel to a diode whose forward current is $I_{\rm d}$. Figure~\ref{fig:PV-equ} shows the equivalent circuit of the PV panel~\cite{a190923.01, a180802.01}. In reality, the PV panel model should add a shunt resistor $R_{\rm sh}$ and a series resistor $R_{\rm s}$. Therefore, according to Kirchhoff's law, the current-voltage~(I-V) characteristics of a PV panel can be described as~\cite{a180802.01,a190923.01}:
\begin{align}
	I_{\rm pv,o}&=I_{\rm ph}-I_{\rm d}-\frac{V_{\rm d}}{R_{\rm sh}},
	\label{equ:I-V-PV}\\
	I_{\rm d}&=I_{\rm 0}({\rm e}^{\frac{V_{\rm d}}{n_{\rm s}{n}V_{T}}}-1),\\
V_{\rm d}&=V_{\rm pv,o}+I_{\rm pv,o}R_{\rm s},
\end{align}
where $I_{\rm 0}$ is the reverse saturation current, and $n_{\rm s}$ is the number of cells connected in series in the PV panel. Thus, for a single-cell PV panel, $n_{\rm s}=1$. In addition, $V_{T}$ is the junction thermal voltage of the diode that related to the temperature:
\begin{equation}
V_{T}=\frac{kT}{q},
\end{equation}
where $n$ is the diode ideality factor, $k$ is the Boltzmann's constant, and $T$ is the panel's temperature in Kelvin.

The photocurrent $I_{\rm ph}$ is related to the light power on the PV panel. We denote $P_{\rm pv,i}$ as the input light power. Then, the photocurrent $I_{\rm ph}$ can be estimated by~\cite{a180720.03}:
\begin{align}
I_{\rm ph}&=\rho P_{\rm pv,i} \nonumber\\
 &=\rho \left\{
\eta_{\rm s}f(d) \left\{
\frac{hc}{q\lambda}\eta_{\rm e}\left[
I_{\rm in}-I_{\rm th}
\right]
\right\}+C
\right\},
\label{equ:Iph-Ppvi}
\end{align}
where $\rho$ is the optical-to-electrical conversion responsivity in~A/W and can be measured under the standard test condition (STC, $25~^\circ$C temperature and $1000$~W/m$^2$ irradiance). From~\eqref{equ:Iph-Ppvi}, we prove the linear relationship between the photocurrent $I_{\rm ph}$ and the light power $P_{\rm pv,i}$.

\begin{figure}[t]
	\centering
	\includegraphics[width=4.3in]{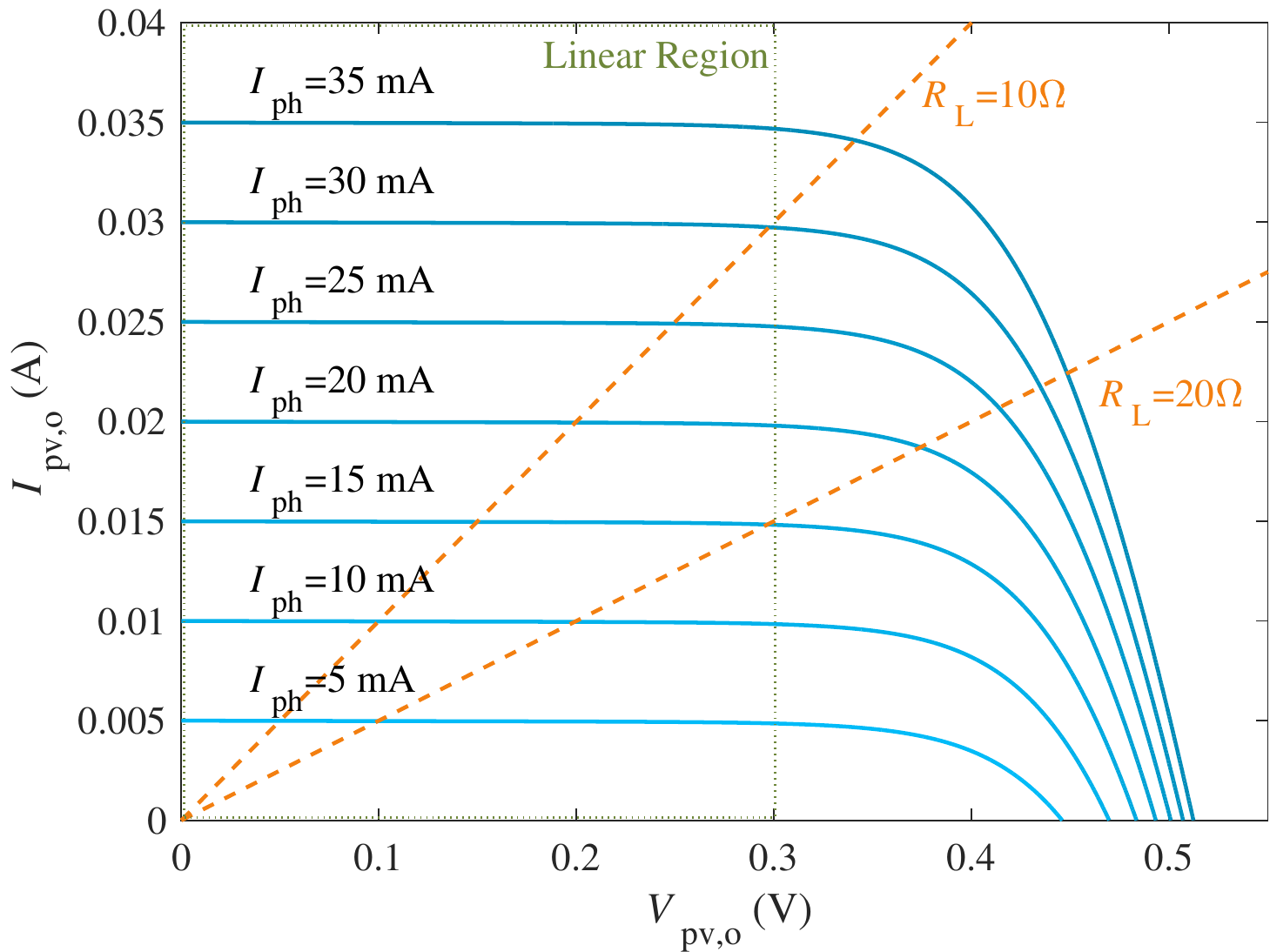}
	\caption{PV panel output current vs. output voltage}
	\label{fig:V-I-PV}
\end{figure}

Figure~\ref{fig:V-I-PV} shows the I-V characteristic curves with different light power. The PV panel operates at different voltage with different load resistance $R_{\rm L}$ although the light power, as well as $I_{\rm ph}$, is at the same level. We can find a linear region in which the PV panel works similarly to a constant-current source regardless of the load resistance $R_{\rm L}$. In the linear region, for each given $P_{\rm pv,i}$, the output current, $I_{\rm pv,o}$, of the PV decreases very slowly with the increasing of output voltage, i.e., $I_{\rm pv,o} \approx I_{\rm ph}$. The load is modeled as an adjustable resistor. We can adjust the load resistance to keep the operating point of the PV panel in its linear region. Note that decreasing the series resistance $R_{\rm s}$ can make the linear region area bigger. However, decreasing the shunt resistance $R_{\rm sh}$ will weaken the constant-current property in the linear region~\cite{a190923.01}.

\subsection{Information Transfer}

As analyzed above, if the temperature shifting is slow, $\eta_{\rm e}$ can be assumed to be a constant. Moreover, the value of $f(d)$ at a specific distance $d$ can be considered as a constant attenuation $\eta_{d}$. From~\eqref{equ:P-I-LD}, \eqref{equ:P-laser} and \eqref{equ:Iph-Ppvi}, the photocurrent of the PV panel is obtained as:
\begin{equation}
\begin{aligned}
 I_{\rm ph}&=\rho \left\{
	\eta_{\rm s}\eta_{d} \left\{
		\frac{hc}{q\lambda}\eta_{\rm e}\left[
			I_{\rm in}-I_{\rm th}
			\right]
		\right\}+C
	\right\}  \\
		  &=\gamma [I_{\rm in}-I_{\rm th}]+\beta, 
\end{aligned}
\label{equ:Iph-Iin}
\end{equation}
where
\begin{equation}
	\gamma=\rho\eta_{\rm s}\eta_{d}\eta_{\rm e}\frac{hc}{q\lambda},~\mbox{and} ~\beta=\rho C.
	\label{equ:gamma}
\end{equation}
In \eqref{equ:Iph-Iin}, $\gamma$ and $\beta$ are constant factors, which verifies the linear modulation ability of the RBCom system.

In the following, we present the modulation model, the transmission loss, the receiver's equivalent circuit, the operating point, the signal power, and the noise impacts.

\subsubsection{Modulation} The input current $I_{\rm in}$ of the PV is modulated to transfer information. The source signal $i_{\rm sig}(t)$ is biased by $I_{\rm bias}$. Therefore, the input current of the pump LD can be described as:
\begin{equation}
\label{equ:Iin=Ibias+Isig}
I_{\rm in}(t)=I_{\rm bias}+i_{\rm sig}(t).
\end{equation}	
Note that $I_{\rm in}$ is the combination of a DC component~$I_{\rm bias}$ and an alternating current (AC) component~$i_{\rm sig}$. The bias current holds up the LD's input current to exceed the threshold $I_{\rm th}$. 

Orthogonal frequency division multiplexing (OFDM) is adopted in modulation, as the frequency response of the PV-ECH circuit is non-flat, which will be detailed later. The signal current $i_{\rm sig}$ is described as:
\begin{equation}
i_{\rm sig}(t)=\sum_{i=0}^{N-1}B_i\cos(\omega_i t + \varphi_i),
\label{equ:isig}
\end{equation}
where $N$ is the number of  subcarriers; $B_i$, $\omega_i$, and $\varphi_i$ are the amplitude,  frequency, phase of the $i$-th subcarrier, respectively.

\begin{figure}[t]
	\centering
	\includegraphics[width=3.5in]{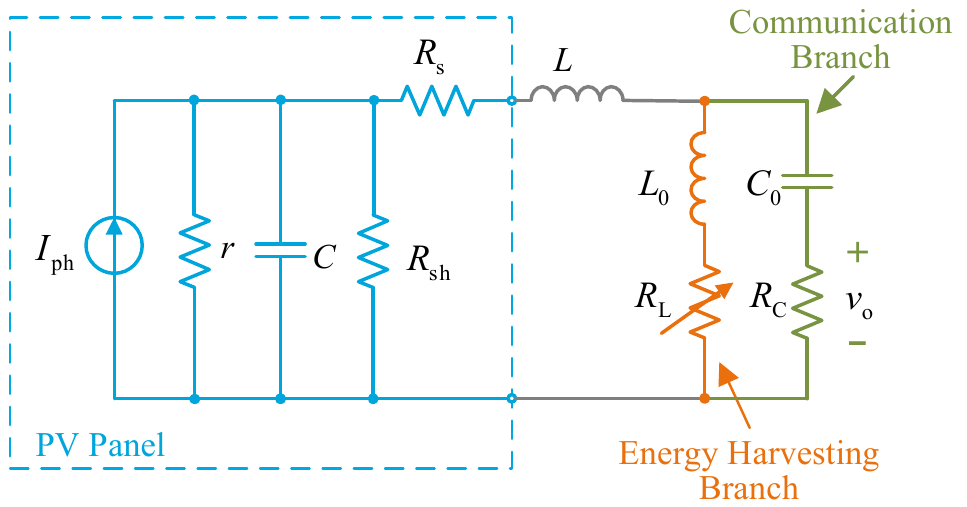}
	\caption{Receiver AC equivalent circuit}
	\label{fig:CEH}
\end{figure}

\subsubsection{Transmission Loss}

	\eqref{equ:fd} illustrates that the transmission loss varies with the distance between the transmitter and the receiver in the RBCom system~\cite{a180820.07}. The transmission loss profile of the RBCom system is distinct from that of other FSO systems. In the RBCom system, diffraction loss is the dominant loss factor. However, in other FSO systems, the dominant factors to cause attenuation are the beam divergence and alignment accuracy [24]. For dozens-of-meter transmission, the resonant beam attenuation can be kept in a low level if the diffraction loss of the cavity is limited through better system design and manufacturing. When choosing the operating wavelength of the RBCom system, we consider visibility, diffraction loss, air absorption/scattering, gain medium material, and PV panel material. Infrared light is better than visible light, as it does not impact people's sight. A small wavelength can provide low diffraction loss. However, small wavelength beam is easier to be absorbed in the atmosphere, so that the wavelength should be as large as possible and should be away from any strong absorption wavelength in the atmosphere. Thus, we should balance the above two factors when choosing wavelength. In addition, the chosen wavelength should ensure that the conversion efficiency of the gain medium and the PV panel is high enough.	

\subsubsection{Receiver's AC Equivalent Circuit}
	Figure~\ref{fig:CEH} depicts the equivalent circuit of the PV panel and CEH module for AC signal transmission~\cite{a180719.01}. In the communication branch, the AC component of the PV's output current is coupled by the capacitor $C_{0}$ and sampled by the communication resistor $R_{\rm C}$. The energy harvesting branch receives the DC component of the PV's output current and chokes the AC component by the inductor $L_{0}$. The operating point is determined by the DC current and the load resistor $R_{\rm L}$. The diode in Fig.~\ref{fig:PV-equ} is modeled as a small-signal equivalent circuit, in which the small-signal dynamic resistance $r$ and the cell capacitance $C$  are connected in parallel. Here $C$ is the shunt capacitance of a transition capacitance $C_{\rm T}$ and a diffusion capacitance $C_{\rm d}$, i.e., $C=C_{\rm T}+C_{\rm d}$. The inductor $L$ models the inductance of wires connected to the PV panel.

	The cell transition capacitance depends on the applied voltage $V_{\rm d}$  at the diode, which is determined by the following equation~\cite{a180808.06}:
	\begin{equation}
		C_{\rm T}(V_{\rm d})={A_{\rm pv}\sqrt{\dfrac{e\epsilon\epsilon_{0}N_{\rm B}}{2({\sqrt{V_{0}-V_{\rm d}}})}}},
	\end{equation}
	where $A_{\rm pv}$ is the area of the PV panel, $\epsilon$ is the permittivity of the semiconductor material~(Si:~$\epsilon=11.7$, GaAs:~$\epsilon=13.1$), $\epsilon_{0}=8.85\times 10^{-12}$ is the vacuum permittivity, $N_{\rm B}$ is the doping concentration in the base region, $V_0$ is the built-in voltage related to the temperature and the doping concentration of the semiconductor~\cite{a180808.01}, and $V_{\rm d}$ is the applied voltage at the diode.
	
	The cell diffusion capacitance also depends on the applied voltage $V_{\rm d}$  at the diode, which is determined by the following equation~\cite{a180808.06, a180808.09}:
	\begin{equation}
		C_{\rm d}(V_{\rm d})=\frac{\tau}{2nV_{T}} I_0{\mathrm e}^{\frac{V_{\rm d}}{nV_{T}}},~\mbox{for}~\omega\tau\ll 1,
	\end{equation}
	where $\omega$ is the angular frequency of the signal; and $\tau$ is the minority carrier lifetime which is a temperature-dependent parameter and can be measured in experiments~\cite{a180808.02}. 
	
	The cell dynamic resistance $r$ relies on the applied voltage $V_{\rm d}$ at the diode, which is determined by the following equation~\cite{Boylestad2008}:
	\begin{equation}
		r(V_{\rm d})=\frac{{\rm d}I_{\rm d}}{{\rm d}V_{\rm d}}=nV_{T}/\left( I_0{\mathrm e}^{\frac{V_{\rm d}}{nV_{T}}}  \right).
		\label{equ:1/r}
	\end{equation}
	  If $I_{\rm d} \gg I_{0}$, \eqref{equ:1/r} can be rewritten as	$r(I_{\rm d})=nV_{T}/I_{\rm d}$.
	 
\subsubsection{Operating Point}
In Fig.~\ref{fig:V-I-PV}, the dashed line represents different load resistances $R_{\rm L}$. The intersection point of one I-V curve and one $R_{\rm L}$ line specifies the operating point. We can find that with a fixed $R_{\rm L}$, the operating point varies with $I_{\rm ph}$. With different operating point, the PV panel achieves different AC parameters, since the cell capacitance and the dynamic resistance depend on the cell voltage.	
		
\subsubsection{Received Signal}

From~\eqref{equ:Iph-Iin}, and \eqref{equ:Iin=Ibias+Isig}, the AC component of the photocurrent is obtained as:
\begin{equation}
	i_{\rm ph,sig}(t)={\gamma}i_{\rm sig}(t).
	\label{equ:iphsig}
\end{equation}
The frequency response of the PV-CEH network, whose input is the photocurrent generated by the PV panel and output is the voltage at the resistor $R_{\rm C}$, is evaluated by~\cite{a180719.01}:

\begin{align}
	H_{\rm ph}(\omega) &=\frac{V_{\rm o}(\omega)}{I_{\rm ph}(\omega)} \nonumber\\
	&=
		\dfrac
			{\dfrac
				{R_{\rm LC}}
				{R_{\rm s}+j{\omega}L+R_{\rm LC}}
			\dfrac
				{R_{\rm C}}
				{\frac{1}{j{\omega}C_{\rm 0}}+R_{\rm C}}
			}
			{
				\dfrac{1}{r}
				+\dfrac{1}{\frac{1}{j{\omega}C}}
				+\dfrac{1}{R_{\rm sh}}
				+\dfrac{1}{R_{\rm s}+j{\omega}L+R_{\rm LC}}
			}
	,
\end{align}
where $\omega$ is the angular frequency, $j:=\sqrt{-1}$, and $R_{\rm LC}$ is the resistance of the parallel network of the energy harvesting branch and the communication branch which is formulated detailedly in~\cite{a180719.01}. According to \eqref{equ:iphsig}, the output signal of the PV-CEH circuit is represented by:
\begin{align}
V_{\rm o}(\omega)&=	\gamma \mathfrak{F}\{ i_{\rm sig}(t) \} H_{\rm ph}(\omega),
\end{align}
where $\mathfrak{F}\{\cdot\}$ denotes the Fourier transform operation.

\begin{table*} 
	\normalsize
	\caption{Parameter List}
	\renewcommand{\arraystretch}{1}
	\centering
	\setlength{\tabcolsep}{4.4mm}
	\begin{tabular}{ l l l l }
		\hline
		Parameter & Symbol &  Value & Unit \\
		\hline
		Reverse saturation current	& $I_{\rm 0}$ & $9.381\times 10^{-9}$ & A \\
		Diode ideality factor		& $n$ & $1.318$ & - \\
		Temperature 				& $T$ & $298.15$ & K \\
		Series resistance 			& $R_{\rm s}$ & $1.3$ & $\Omega$ \\
		Shunt resistance			& $R_{\rm sh}$ & $5000$ & $\Omega$ \\
		Conversion responsivity 	& $\rho$ & $0.746$ &  A/W \\
		Dynamic resistance			& $r$ & $839.5$ & $\Omega$\\
		Cell capacitance			& $C$ & $26.6$  & nF 	\\

		AC choke inductor			& $L_{\rm 0}$ & $40$ & mH \\
		AC coupling capacitor 		& $C_{\rm 0}$ & $6$ & pF \\

		Load resistor    		& $R_{\rm L}$  &  $0.6$  &  $\Omega$\\	
		
		Optical efficiency of Rx antenna & $\eta_{\rm Rx}$ & $0.5$ & - \\
		Background irradiance 		& $H_{\rm bkg}$ & $0.2$ & W$\cdot$m$^{-2}\cdot$nm$^{-1}\cdot$sr$^{-1}$ \\
		Bandwidth of optical filter & $B_{\rm IF}$ & $20$ & nm \\
		Solid angle of view of PV	& $\Phi_{\rm rcv}$ & $2\uppi(1-\cos30^\circ$) & sr \\
		Transmittance of mirror R2	& $\Gamma$ & $0.05$ & - \\	
		
		\hline
	\end{tabular}
	\label{tab:params}
\end{table*}

\subsubsection{Receiver Noise}

The noises at the receiver include the shot noise and the thermal noise of the circuit. The shot noise is related to the received light power, including the laser power and the background radiance~\cite{a180719.01,a180718.01}.

The background radiance in the environment comes from the sunlight or the lamplight. The  background radiance power received by the PV panel is formulated as~\cite{a180718.01}:
\begin{equation}
	P_{\rm bkg}={\eta}_{\rm Rx}H_{
	\rm bkg}{B}_{\rm IF}A_{\rm Rx}\Phi_{\rm Rx}\Gamma,
\end{equation}
where $\eta_{\rm Rx}$ is the optical efficiency of the Rx antenna; $H_{\rm bkg}$ is the background irradiance in~W$\cdot$m$^{-2}$$\cdot$nm$^{-1}$$\cdot$sr$^{-1}$; $B_{\rm IF}$ is the optical bandwidth of the filter installed behind R2; $A_{\rm Rx}$ is the receiving area of the Rx antenna; and $\Phi_{\rm Rx}$ is the solid angle of Rx antenna's field of view. The background radiance in the RBCom system passes through the mirror R2 which has a low transmittance $\Gamma$. Therefore, the background radiance experiences an additional attenuation that  general laser communication systems or VLC systems can not provide. From \eqref{equ:Iph-Ppvi}, the photocurrent caused by the background radiance is obtained as:
\begin{equation}
	i_{\rm ph,bkg}={\rho}P_{\rm bkg}.
\end{equation}

The shot noise comes from the photoelectric conversion in the PV panel and can be modeled  by a Gaussian distribution which has a flat power spectral density~(PSD). The one-side PSD of the shot noise in~A$^2$/Hz can be obtained by~\cite{a180730.03}:
\begin{equation}
N_{\rm pv,{\rm Hz}^{-1}}^{\rm sh}=2q\rho{P_{\rm pv,i}},
\end{equation}
where $q$ is the electron charge; and $P_{\rm pv,i}$ is the average optical power over the PV panel which comprises the laser power $P_{\rm laser}$ and the background radiance power $P_{\rm bkg}$. The shot noise shares the same network, i.e., the same frequency response, with the signal. Therefore, the one-side PSD of the shot noise expressed in V$^2$/Hz output from $R_{\rm C}$ is obtained as:
\begin{equation}
N_{\rm o}^{\rm sh}(\omega)=	\left|H_{\rm ph}(\omega)\right|^2 \times 2q{\rho}({P_{\rm laser}+P_{\rm bkg}}).
\end{equation}

The thermal noises are generated by the resistors in the equivalent circuit depicted in Fig.~\ref{fig:CEH}. The one-side PSD in A$^2$/Hz of the thermal noise generated by the resistor $R_{\rm C}$ can be obtained as~\cite{a180730.03}:
\begin{equation}
N_{\rm o,RC}^{\rm th}=4kTR_{\rm C},
\end{equation}
where $k$ is Boltzmann's constant, $T$ is the temperature of $R_{\rm C}$ in Kelvin. Likewise, the one-side PSD in V$^2$/Hz of the total output noise contributed by all the thermal noises in the circuit can be obtained as~\cite{a180719.01}:
\begin{align}
N_{\rm o}^{\rm th}(\omega)
& =|H_{\rm RC}(\omega)|^2 N_{\rm o,RC}^{\rm th}
  +|H_{\rm Rsh}(\omega)|^2 N_{\rm o,sh}^{\rm th} \nonumber\\
&~~~+|H_{\rm RL}(\omega)|^2 N_{\rm o,RL}^{\rm th}
  +|H_{\rm r}(\omega)|^2 N_{\rm o,r}^{\rm th} \nonumber\\
&~~~+|H_{\rm Rs}(\omega)|^2 N_{\rm o,Rs}^{\rm th},
\end{align}
where $N_{\rm o,sh}^{\rm th}$, $N_{\rm o,RL}^{\rm th}$, $N_{\rm o,r}^{\rm th}$, and $N_{\rm o,Rs}^{\rm th}$ are the one-side PSD of the thermal noises that the resistors $R_{\rm sh}$, $R_{\rm L}$, $r$, and $R_{\rm s}$ generate, respectively; and $|H_{\rm RC}(\omega)|^2$, $|H_{\rm Rsh}(\omega)|^2$, $|H_{\rm RL}(\omega)|^2$, $|H_{\rm r}(\omega)|^2$, and $|H_{\rm Rs}(\omega)|^2$ are the frequency responses in V$^2$/A$^2$ of the networks that the thermal noises from  $R_{\rm C}$, $R_{\rm sh}$, $R_{\rm L}$, $r$, and  $R_{\rm s}$ experience, respectively.

\subsubsection{Signal-to-Noise Ratio}  
Because the PV-CEH circuit exhibits a non-flat frequency response, OFDM is recommended to be adopted in modulation. Adopting OFDM in such a frequency-selective channel is a significant way to maximize the utilization of the channel capacity, as  introduced in~\cite{a191213.01,a191213.02}. Using OFDM, the frequency selective channel is separated into many subchannels. In each subchannel, modulation scheme such as phase shift keying (PSK) and quadrature amplitude modulation (QAM) is adopted. Different modulation scheme is adopted in each subchannel to adapt this subchannel's charateristic, so the performance of the whole channel can be maximized. The signal-to-noise ratio (SNR) of each subchannel can be estimated by:
\begin{equation}
{\rm SNR}_i= \frac{\gamma^2 \left(\sigma_s^2/N\right) \left|H_{\rm ph}(\omega_i)\right|^2}
        { w \left[
        	{N_{\rm o}^{\rm sh}(\omega_i)}
		   +{N_{\rm o}^{\rm th}(\omega_i)}
		  \right] 
	  },
\end{equation}
where $\sigma_{\rm s}^2$ is the variance of the time-domain signal $i_{\rm sig}$, $N$ is the number of subcarriers, and $w$ is the bandwidth of each subchannel.

\section{Numerical Results}
\label{sec:res}

In this section, we at first introduce the parameters used in the following computation. Next, we investigate the operating point and the dynamic parameters of the PV panel. Then, we depict the network frequency response profiles of the signal and each noise source. At last, we investigate the SNR at the receiver and the communication performance.
\begin{figure}
	\centering
	\includegraphics[width=4.3in]{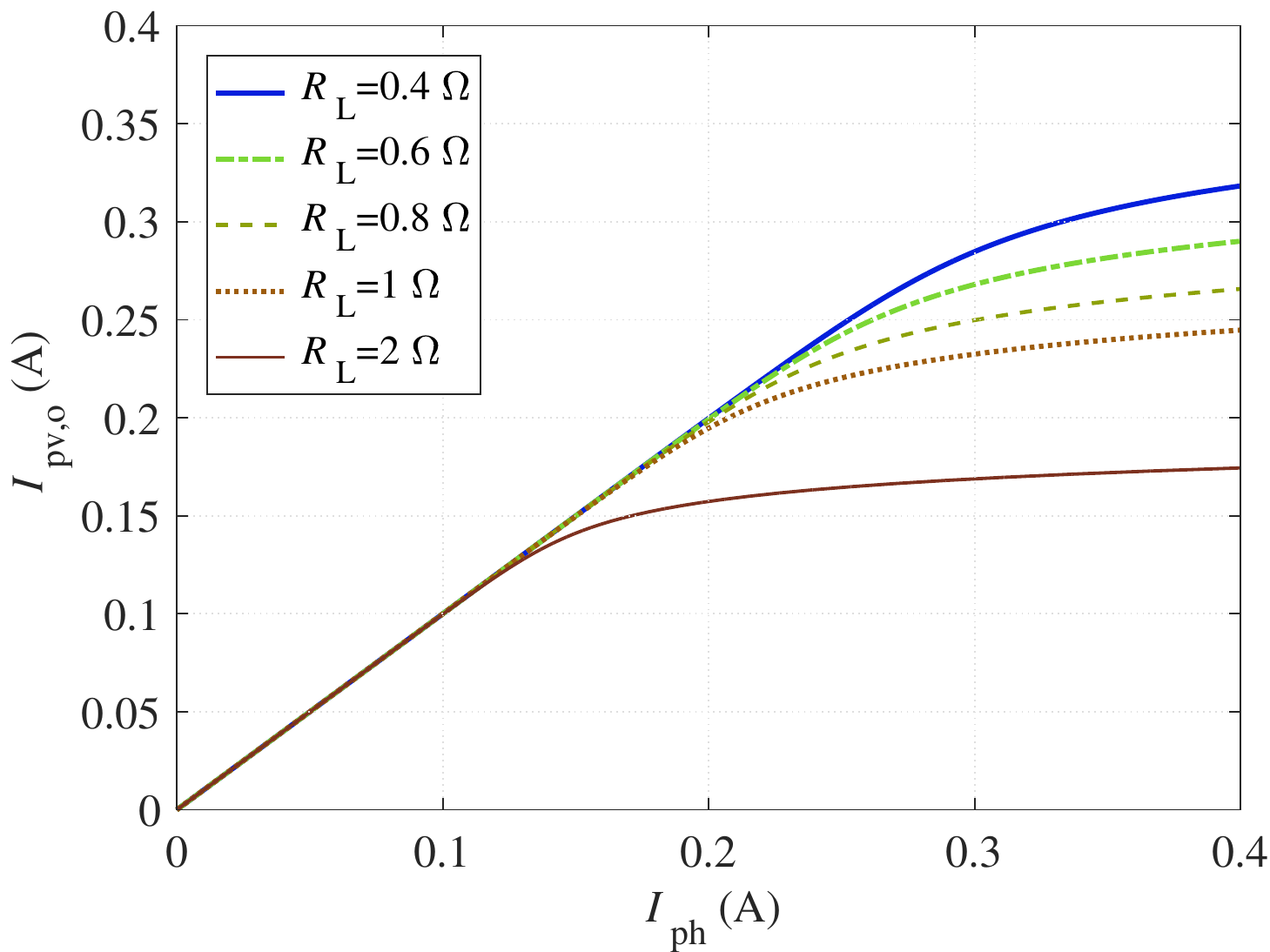}
	\caption{PV panel output current $I_{\rm pv,o}$ \textit{vs.} photocurrent $I_{\rm ph}$}
	\label{fig:Ipvo-Iph}
\end{figure}

\subsection{Parameters}

The default parameters adopted in the following computation are listed in Table~\ref{tab:params}, unless otherwise specified. For simplicity, we assume using a single-cell PV panel, i.e. cell number $n_{\rm s}=1$. The area of the PV panel $A_{\rm pv}$ and the area of the Rx antenna $A_{\rm Rx}$ are assumed to be 1~cm$^2$.  The capacitance-voltage~(C-V) characteristic curve of the PV panel are obtained from~\cite{a180811.08}. The parameters of the PV cell $I_{0},n,T,R_{\rm s}, R_{\rm sh}$ are obtained by searching the optimal values to make the model expressed by~\eqref{equ:I-V-PV} approximate to the I-V curve. As analyzed in~\cite{a180817.01}, the maximum photoelectric conversion efficiency of a mono-Si PV cell can reach to $25\%$ under STC. With this efficiency, we obtained the conversion responsivity $\rho=746$~mA/W. The parameters of background irradiance $H_{\rm bkg}$ are obtained from~\cite{a180718.01}. 
%
The wire inductance $L=120$~nH is obtained from~\cite{a180719.01}. We also consider the smaller wire inductance, i.e., $L=10$~nH. With different $L$, the communication resistor $R_{\rm C}$ should be adjusted to achieve a flat passband in the frequency response.

According to \cite{a180820.07}, the conversion efficiency from the input electrical power $P_{\rm in}$ of the pump source to the stored power of the gain medium $P_{\rm s}$ is conducted as $42.3\%$, and the conversion efficiency  from $P_{\rm store}$ to the output laser power $P_{\rm laser}$ is measured as $11.45\%$ over $1.675$~m transmission distance. Hence, the transmission efficiency from the input electrical power to the coupled output laser power is $4.84$\%. As stated in \cite{a180720.01}, the external quantum efficiency, $\eta_{\rm e}$, of a semiconductor laser is up to $90\%$. Hence, we assume $\eta_{\rm s}\eta_{d}=5.4\%$ in the following computation though this value will be greater in practice. Besides,  $808$~nm semiconductor laser is employed as the pump source. In this case, 1064 nm coupled output laser from the resonant beam cavity is obtained. From \eqref{equ:gamma}, we obtain $\gamma=0.0557$. The variance of the signal $\sigma_s^2$ is assumed to be $0.01~\mbox{A}^2$. Therefore, the variance of the AC component of the photocurrent is $\sigma_{\rm ph,ac}^2 = 3.1 \times 10^{-5}~\mbox{A}^2$. The average power of the received laser  $P_{\rm laser}$ is assumed to be $200$~mW, which can provide an proper operating point. In this case, the DC component of the photocurrent $I_{\rm ph,dc}$ is $149.2$~mA. 

\begin{figure}
	\centering
	\includegraphics[width=4.3in]{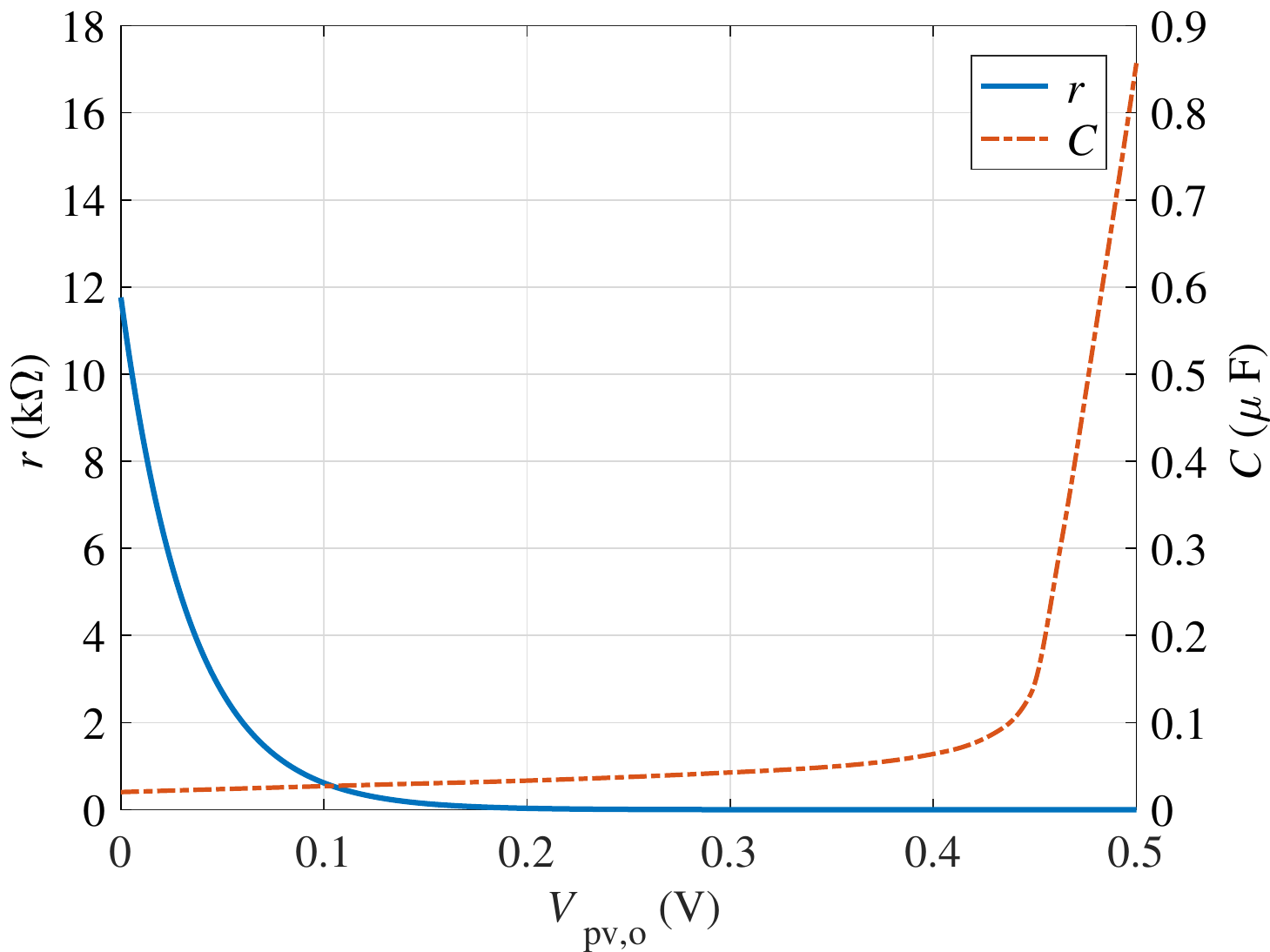}
	\caption{Dynamic resistance $r$ and cell capacitance $C$ \textit{vs.} cell voltage $V_{\rm pv,o}$}
	\label{fig:r-C-Vpvo}
\end{figure}

\subsection{Operating Point}
The value of load resistance $R_{\rm L}$ and photocurrent $I_{\rm ph}$ determine the operating point. Given a certain $R_{\rm L}$, the relationship between the PV output current $I_{\rm pv,o}$ and the photocurrent $I_{\rm ph}$ is obtained, as shown in Fig.~\ref{fig:Ipvo-Iph}. Each curve has two linear segments, between which a bent segment connects them. If the operating point  lies in the bent segment, a high-power (large peak-to-peak current of $I_{\rm ph}$) signal will face with heavy distortion, as $I_{\rm pv,o}$ varies with $I_{\rm ph}$ nonlinearly. However, in the right segment region, $I_{\rm pv,o}$ changes slightly with the variation of $I_{\rm ph}$, which means the amplitude of the output signal will be heavily reduced.

The dynamic parameters depend on the voltage of the PV cell. Hence, they can be determined at a specific operating point. Figure~\ref{fig:r-C-Vpvo} illustrates the variation in the dynamic resistance $r$ and the cell capacitance with the cell voltage $V_{\rm pv,o}$. When the cell voltage is low, $r$ achieves a large value. Yet, when $V_{\rm pv,o}$ is high, $r$ becomes very low. On the contrary, $C$ is at a low level and increases rapidly when $V_{\rm pv,o}$ grows to a high level. From this figure, we can conclude that a low cell voltage is helpful for communication, as a large dynamic resistance and a low cell capacitance can improve the gain of the signal network.

\begin{figure}
	\centering
	\includegraphics[width=4.3in]{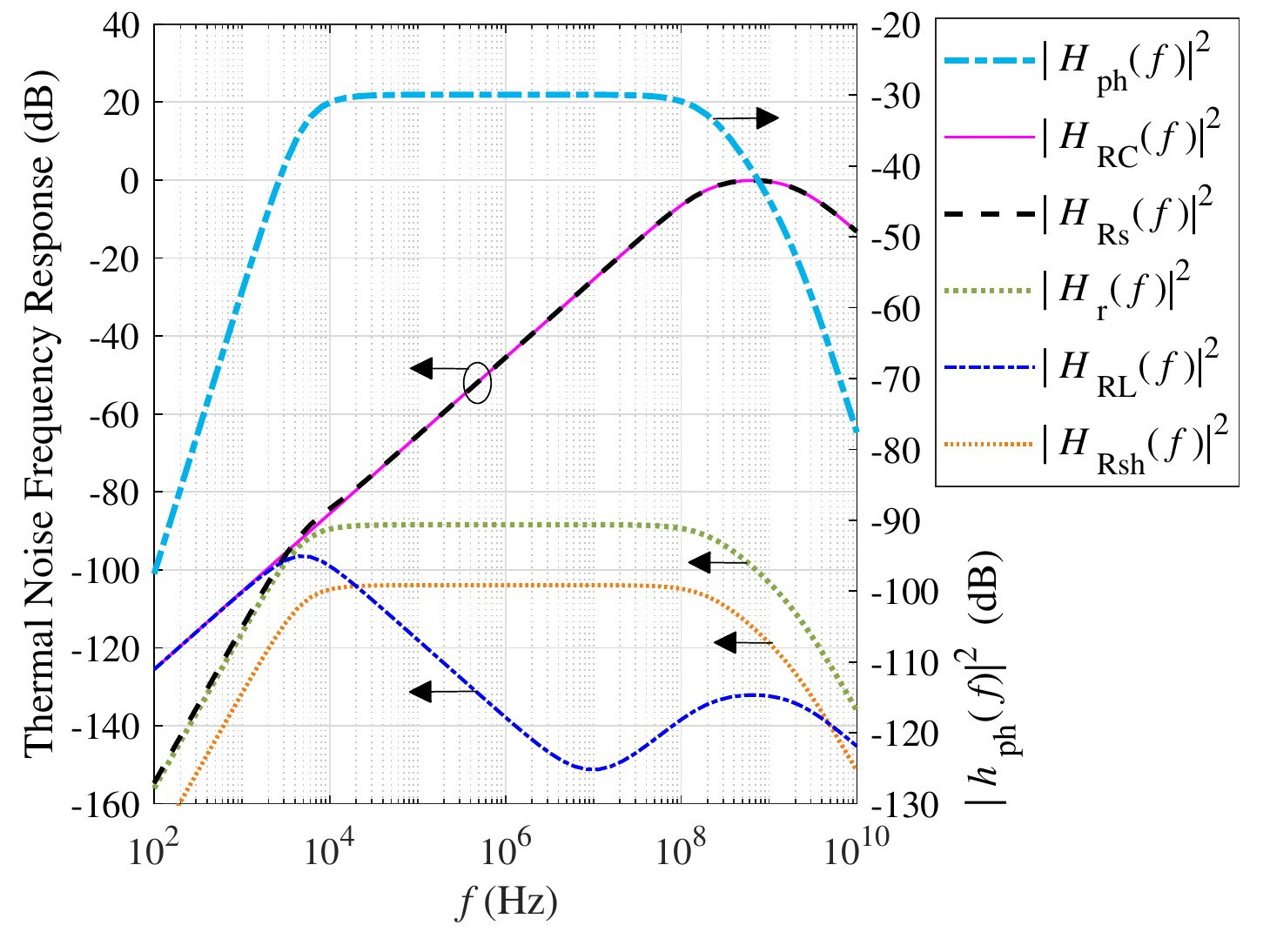}
	\caption{Frequency responses ($L=10~\mbox{nH}$ and $R_{\rm C}=140~\Omega$)}
	\label{fig:fResponse-10nH}
\end{figure}

\begin{figure}
	\centering
	\includegraphics[width=4.3in]{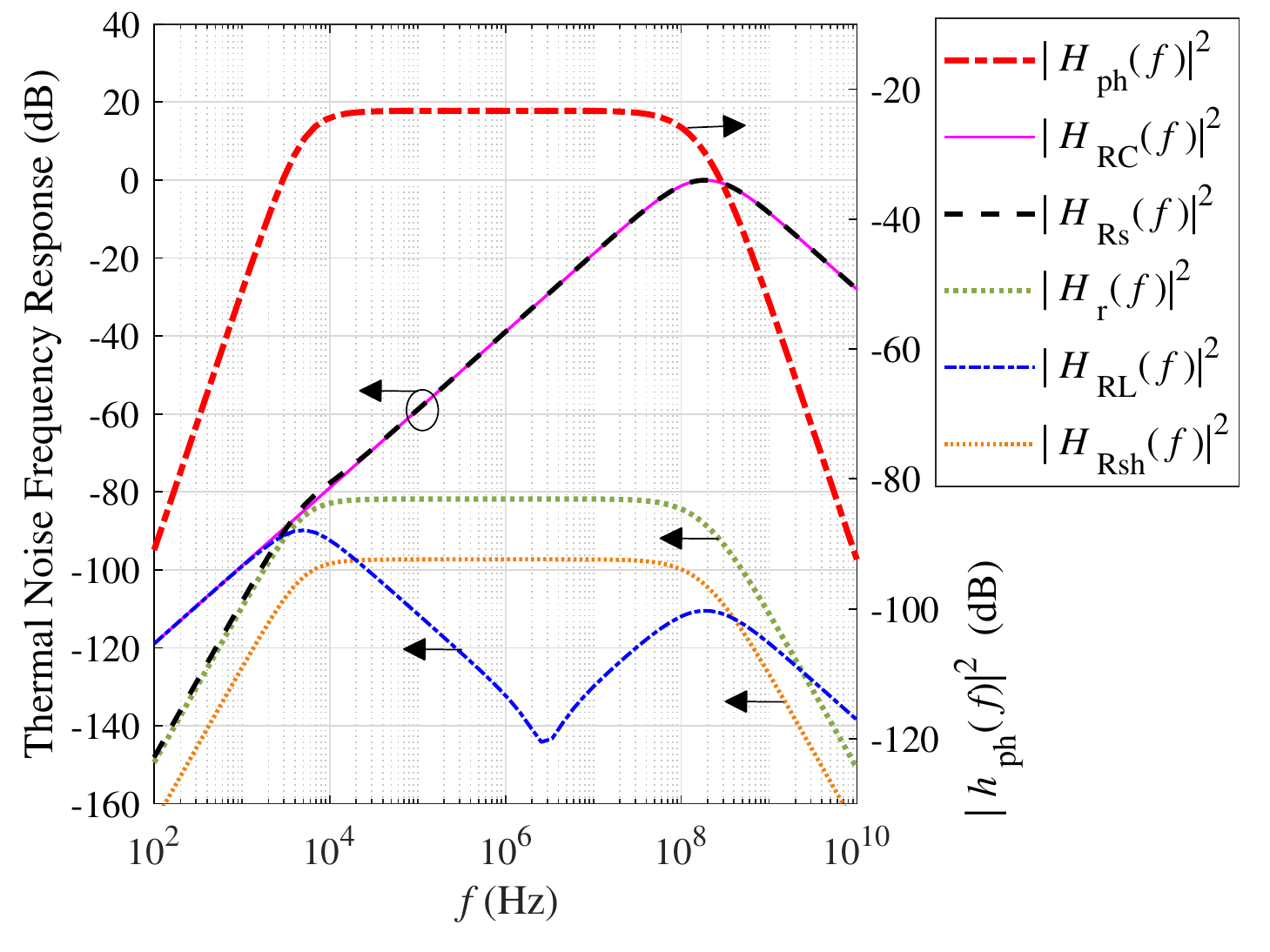}
	\caption{Frequency responses ($L=120~\mbox{nH}$ and $R_{\rm C}=300~\Omega$)}
	\label{fig:fResponse-120nH}
\end{figure}

\subsection{Frequency Response}

In the receiver, there are information signal, shot noise and five thermal noises. Figure ~\ref{fig:fResponse-120nH} and~\ref{fig:fResponse-10nH} depict the frequency responses of the networks that the signal and the noises experience. Two cases, where $\{L,R_{\rm C}\}=\{120~\mbox{nH},300~\Omega\}$ and $\{L,R_{\rm C}\}=\{10~\mbox{nH},140~\Omega\}$, are analyzed. The frequency response, $|H_{\rm ph}(f)|^2$, of the signal network, from the input $I_{\rm ph}$ to the output $v_{\rm o}$, exhibits a high flat gain in the middle frequency region. We find the bandwidths of the signal networks with these two cases approximate to $120$~MHz ($L=120$~nH) and $200$~MHz ($L=10$~nH), respectively. The signal and the shot noise share the same gain, since they pass through the same network. The other five curves in the figures are the frequency responses of the networks that the five thermal noises experience, respectively. The most influential thermal noises come from $R_{\rm C}$ and $R_{\rm s}$.

\subsection{SNR and Channel Capacity}

\begin{figure}
	\centering
	\includegraphics[width=4.4in]{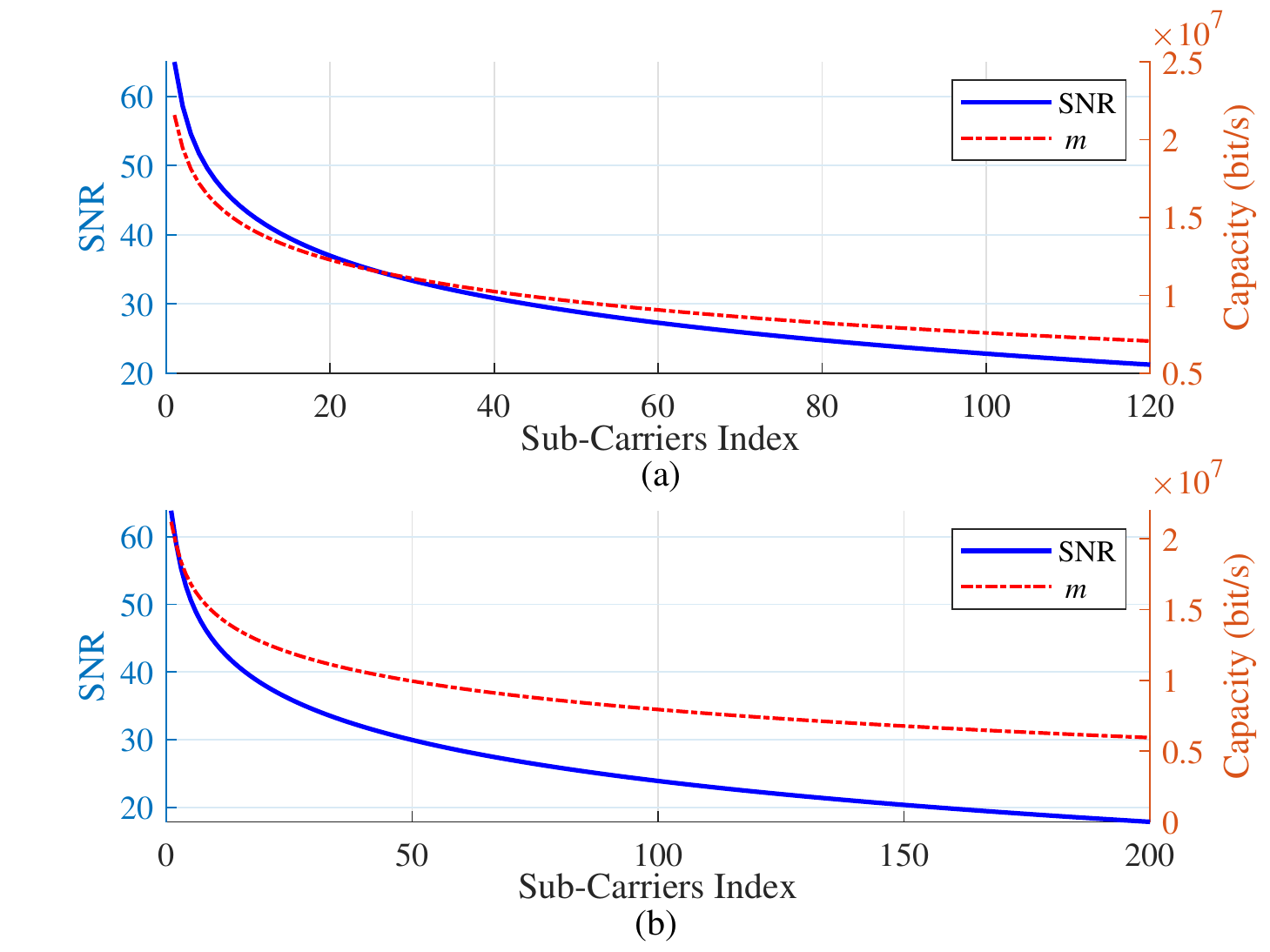}
	\caption{Signal-to-noise ratio and channel capacity of each subcarrier: (a) $L=120~\mbox{nH}$, $R_{\rm C}=300~\Omega$; (b) $L=10~\mbox{nH}$, $R_{\rm C}=140~\Omega$}
	\label{fig:SNR-f}
\end{figure}

The above frequency response profiles are non-flat, especially for thermal noises. Therefore, OFDM is employed to improve the communication performance. We evaluate the SNR and the channel capacity of the RBCom system in the aforementioned two cases.
The bandwidth of each subchannel is $1~\mbox{MHz}$. Hence, the number of subcarriers in these two cases are $120$ ($L=120$~nH) and $200$ ($L=10$~nH), respectively. Figure~\ref{fig:SNR-f} shows the SNR and the channel capacity of each subchannel. The SNR and the channel capacity are high when the frequency of the subcarrier is low. As the subcarrier frequency grows, the SNR and the channel capacity decreases gradually. In summary, the total capacities of the RBCom system in these two cases are $1.19$~Gb/s ($L=120$~nH) and $1.76$~Gb/s ($L=10$~nH), respectively. Besides, smaller $L$ provides higher channel capacity.

\subsection{Output charging power and Channel Capacity}

\begin{figure}
	\centering
	\includegraphics[width=4.4in]{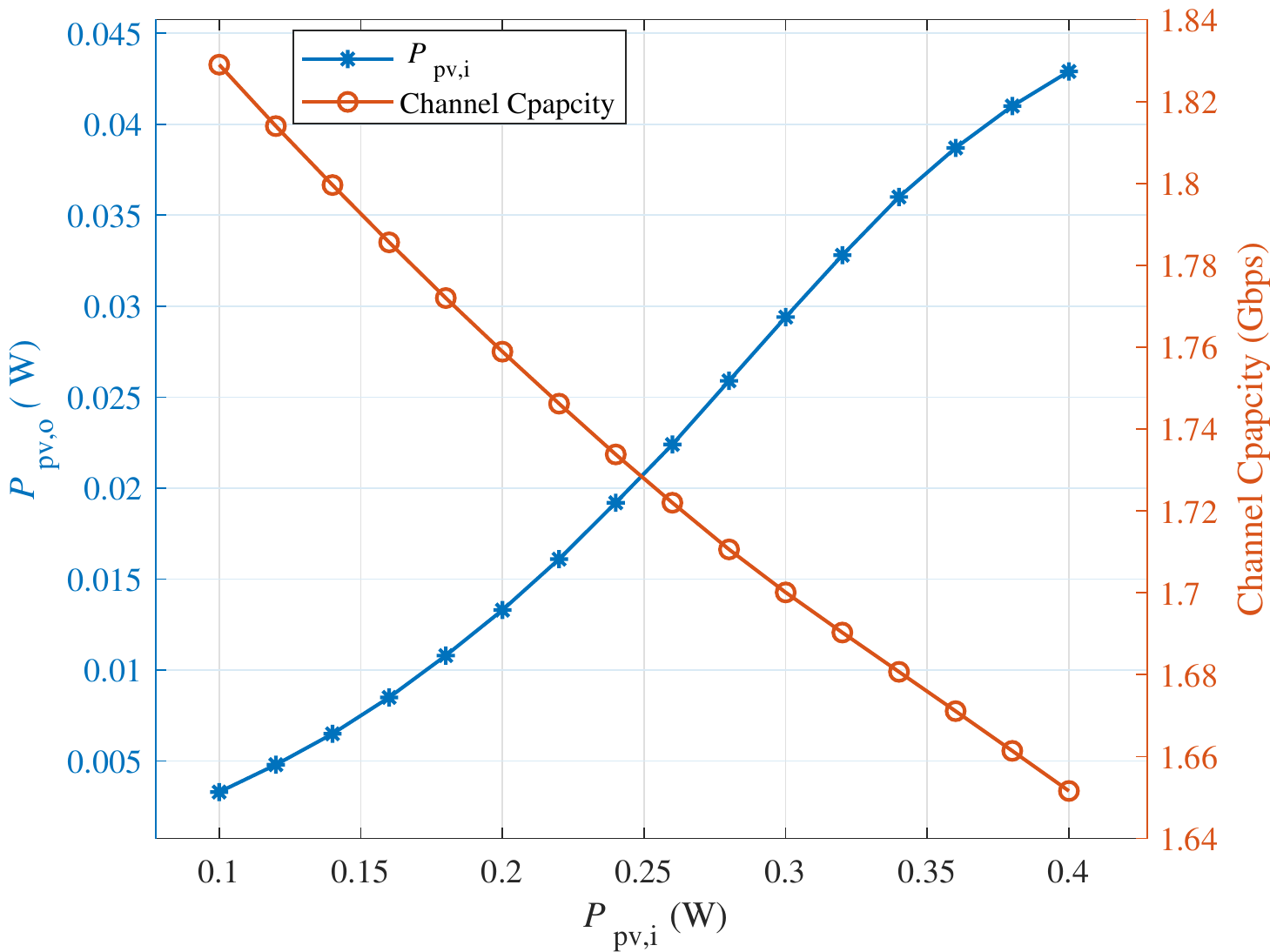}
	\caption{Output charging power and channel capacity ~($L=10~\mbox{nH}$, $R_{\rm C}=140~\Omega$)}
	\label{fig:PoutCapa}
\end{figure}

Figure~\ref{fig:PoutCapa} shows how the output charging power and the channel capacity vary with the input laser power of the PV panel. Here the inductance $L=10~\mbox{nF}$, and the resistor $R_{\rm C}=140~\Omega$. The variance of the signal $\sigma_s^2$ is still assumed to be $0.01~\mbox{A}^2$. As depicted in Fig.~\ref{fig:PoutCapa}, as the input laser power increases, the output charging power $P_{\rm pv,i}$ increases and the channel capacity decreases. Although high input laser power can provide high charging power for the battery, the channel capacity is limited by the high input power, because large photocurrent results in large short noise.  

\subsection{Discussion}

The channel capacity is limited by the bandwidth of the PV panel, as it exhibits large cell capacitance. We use PV panel in the RBCom system to receive not only data but also energy. Therefore, this system can be employed in some special scenarios where energy and data are simultaneously desired. As concluded in \cite{a180820.09}, both wireless charging and data transfer are important aspects in 6G mobile communications. Besides, this manuscript only analyzes the case where only one transmitter sends data to the receiver. In fact, multiple transmitters can be assembled in a base station to transfer data and power to multiple receivers, which can provide an improved transfer capability.


\section{Conclusions}
\label{sec:con}
In this paper, we present the resonant beam communication (RBCom) system for optical data and power transfer with wide coverage and mobility. Relying on the analytical evaluation, we find the RBCom system can achieve more than 40 mW charging power and $1.6$ Gbit/s channel capacity with OFDM scheme. The RBCom system connects the transmitter and the receiver with an optical beam link, which can satisfy the requirements of wireless charging and high-rate communications. 

There are several directions of future research,
including: {1) analyzing parameters like bit error rate (BER), quality of service (QoS), Q factor, link uptime, and eye patterns by experiments; 2) optimizing the RBCom system design parameters to enhance data rate; 3) studying wavelength division multiplexing (WDM) in RBCom system to enhance data rate; and 4) optimizing system design to enhance power transfer capability.


%

\appendices

%



\ifCLASSOPTIONcaptionsoff
  \newpage
\fi



\bibliographystyle{IEEEtran}
%
\bibliography{OWIPT-RBC}
%
%

%

%
%







\end{document}